\PassOptionsToPackage{unicode}{hyperref}
\PassOptionsToPackage{hyphens}{url}
\PassOptionsToPackage{dvipsnames,svgnames,x11names}{xcolor}
\documentclass[
]{article}
\usepackage{xcolor}
\usepackage[margin=1in]{geometry}
\usepackage{amsmath,amssymb}
\usepackage{iftex}
\ifPDFTeX
  \usepackage[T1]{fontenc}
  \usepackage[utf8]{inputenc}
  \usepackage{textcomp} 
\else 
  \usepackage{unicode-math} 
  \defaultfontfeatures{Scale=MatchLowercase}
  \defaultfontfeatures[\rmfamily]{Ligatures=TeX,Scale=1}
\fi
\usepackage{lmodern}
\ifPDFTeX\else
\fi
\IfFileExists{upquote.sty}{\usepackage{upquote}}{}
\IfFileExists{microtype.sty}{
  \usepackage[]{microtype}
  \UseMicrotypeSet[protrusion]{basicmath} 
}{}
\makeatletter
\@ifundefined{KOMAClassName}{
  \IfFileExists{parskip.sty}{%
    \usepackage{parskip}
  }{
    \setlength{\parindent}{0pt}
    \setlength{\parskip}{6pt plus 2pt minus 1pt}}
}{
  \KOMAoptions{parskip=half}}
\makeatother
\providecommand{\tightlist}{%
  \setlength{\itemsep}{0pt}\setlength{\parskip}{0pt}}
\usepackage{float}
\usepackage{graphicx}
\usepackage[ruled,vlined]{algorithm2e}
\usepackage{bookmark}
\IfFileExists{xurl.sty}{\usepackage{xurl}}{} 
\makeatletter
\@ifundefined{xmpquote}{}{}
\makeatother
\hypersetup{
  colorlinks=true,
  linkcolor={Maroon},
  filecolor={Maroon},
  citecolor={Blue},
  urlcolor={Blue},
  pdfcreator={LaTeX via pandoc}}

\author{}
\date{}

\begin{document}

\begin{center}
{\LARGE CAQ: \underline{C}lustered Asymptotically \underline{A}nytime-Valid \underline{Q}uantile Treatment Effects for Online Experiments\par}
\vspace{0.9em}
{\large Diane Lu\textsuperscript{1} and Eric J. Oh\textsuperscript{2}\par}
\vspace{0.4em}
\textsuperscript{1}LaunchDarkly, \texttt{dlu@launchdarkly.com}\\
\textsuperscript{2}LaunchDarkly, \texttt{eoh@launchdarkly.com}
\end{center}

\section{1. Introduction}\label{introduction}

\textbf{Motivation.} Quantile metrics are ubiquitous in online
experimentation, particularly as continuously monitored primary or
guardrail metrics. Consider an e-commerce team continuously monitoring
the 95th percentile of a new checkout flow's latency (the slow requests
that lose customers). They would like to stop an experiment with
degrading latency immediately to minimize impact, rather than waiting
for a preplanned horizon. This setting imposes three technical demands
at once:

\begin{enumerate}
\def\labelenumi{(\arabic{enumi})}
\tightlist
\item
  \textbf{Quantiles} rule out standard mean-based CLT intervals.
\item
  Continuous monitoring requires \textbf{anytime-validity} to prevent
  inflating Type I errors from ``peeking''.
\item
  \textbf{Clustered} data (customers sending multiple correlated
  requests) breaks i.i.d. assumptions, producing overconfident intervals
  that miscover.
\end{enumerate}

\textbf{Contribution.} We introduce \textbf{CAQ} (\textbf{C}lustered,
asymptotically \textbf{A}nytime-valid \textbf{Q}uantile treatment
effect, or QTE): a cluster-robust asymptotic confidence sequence
(AsympCS) for a quantile, valid under any number of looks in the
asymptotic sense of Waudby-Smith et al.~\hyperref[ref6]{{[}6{]}}, and a
sequential test for the QTE between arms. To our knowledge, CAQ is the
first to combine all three axes. Prior work covers only pairs: clustered
quantiles at a fixed horizon {[}\hyperref[ref1]{1}, \hyperref[ref4]{4},
\hyperref[ref5]{5}{]}, sequential quantiles only under i.i.d. sampling
{[}\hyperref[ref2]{2}, \hyperref[ref3]{3}{]}. We benchmark one
representative per gap: Deng et al. \hyperref[ref1]{{[}1{]}}, the
cluster-robust fixed-horizon interval most platforms use, and
Howard-Ramdas \hyperref[ref2]{{[}2{]}}, the standard sequential quantile
confidence sequence (CS), which \hyperref[ref3]{{[}3{]}} also builds on.

\section{2. Setup and Method}\label{sec-method}

\textbf{Setup.} Randomization is at the level of a unit (a user) \(i\),
a cluster of correlated events; we use \emph{unit} and \emph{cluster}
interchangeably. Unit \(i\) contributes \(N_i\) within-unit events
\(\{X_{ij} : j = 1, \dots, N_i\}\) (e.g.~its requests) over a fixed
window after first exposure. Only window-completed units enter the
estimate, so units enter as i.i.d. draws; \(K\) counts them (the running
sample size, in units, not events). Let \(Q(p) = F^{-1}(p)\) be the true
\(p\)-quantile of the event distribution \(F\) and
\(\hat\theta_p = X_{(\lceil np\rceil)}\) the pooled sample quantile over
the \(n = \sum_i N_i\) events seen. Two independent arms \(c, t\) give
the estimand \(\Delta_p = Q_t(p) - Q_c(p)\), estimated by
\(\hat\Delta = \hat\theta_p^{(t)} - \hat\theta_p^{(c)}\); we test
\(H_0 : \Delta_p = 0\).

\textbf{One-arm.} Let \(S_i = \#\{j : X_{ij} \le Q(p)\}\). Within a unit
events are correlated, but the pairs \((S_i, N_i)\) are i.i.d. across
units, so the clustered CDF estimate \(\hat p = \bar S / \bar N\) is a
ratio of i.i.d. per-unit averages targeting \(\mu_S/\mu_N = p\). Using a
first-order Taylor expansion, we show the clustered CDF estimate
\(\hat p\) is an asymptotically linear estimator. It therefore inherits
a \((1-\alpha)\)-AsympCS \hyperref[ref6]{{[}6{]}}, which we invert into
a quantile confidence sequence (Algorithm \ref{alg:onearm}) via order
statistics (full derivation in \hyperref[app-a]{Appendix A}). The
order-statistic gap is wide where events are sparse and narrow where
they are dense, so it reflects the local density without needing to
estimate it.

\textbf{Two-arm.} The two-arm QTE test (Algorithm \ref{alg:twoarm}) runs
one AsympCS per arm, reads the bounding factor at the \emph{total} count
\(m = K_c + K_t\), and combines the per-arm quantile-space half-widths
as \(\hat\Delta \pm \sqrt{h_c^2 + h_t^2}\). At its core, each per-arm
quantile estimator is asymptotically linear, and inverse-probability
weighting fuses the two into a single such estimator of the QTE over the
shared count \(m\), which inherits a \((1-\alpha)\)-AsympCS
\hyperref[ref6]{{[}6{]}}. See the full derivation in
\hyperref[app-a]{Appendix A}.

\begin{algorithm}[H]
\DontPrintSemicolon
\KwIn{quantile level $p$, level $\alpha$, tuning $\rho$; units arriving sequentially}
\KwOut{asymptotically anytime-valid confidence sequence for the quantile $Q(p)$}
\ForEach{look after $K$ units \textnormal{(pooled size $n \gets \sum_{i=1}^{K} N_i$)}}{
  $\hat\theta_p \gets X_{(\lceil np\rceil)}$;\ \ $S_i \gets \#\{j : X_{ij} \le \hat\theta_p\}$;\ \ $\hat p \gets \bar S / \bar N$\;
  $\hat V \gets \bar N^{-2}\big(\hat\sigma_S^2 - 2\hat p\,\hat\sigma_{SN} + \hat p^2\hat\sigma_N^2\big)$;\ \ $\beta_K \gets b(K)\sqrt{\hat V / K}$, \ $b(K)=\sqrt{\tfrac{2(K\rho^2+1)}{K\rho^2}\log\tfrac{\sqrt{K\rho^2+1}}{\alpha}}$\;
  $L \gets \lceil n(p-\beta_K)\rceil$,\ \ $U \gets \lfloor n(p+\beta_K)\rfloor + 1$;\ \ \textbf{report} $[\,X_{(L)},\, X_{(U)}\,]$\;
}
\caption{One-arm, clustered, asymptotically anytime-valid quantile confidence sequence}
\label{alg:onearm}
\end{algorithm}

\begin{algorithm}[H]
\DontPrintSemicolon
\KwIn{level $\alpha$, tuning $\rho$; two independent arms $c, t$ arriving sequentially}
\KwOut{asymptotically anytime-valid test / CS for the QTE $\Delta_p = Q_t(p) - Q_c(p)$}
\ForEach{look with $K_c, K_t$ units per arm \textnormal{($m \gets K_c + K_t$)}}{
  \ForEach{arm $a \in \{c, t\}$}{
    compute $\hat\theta_p^{(a)}, \hat V_a$ as in Algorithm~\ref{alg:onearm};\ \ $\beta_a \gets b(m)\sqrt{\hat V_a / K_a}$\;
    $L_a \gets \lceil n^{(a)}(p-\beta_a)\rceil$,\ \ $U_a \gets \lfloor n^{(a)}(p+\beta_a)\rfloor + 1$;\ \ $h_a \gets \tfrac12\big(X^{(a)}_{(U_a)} - X^{(a)}_{(L_a)}\big)$\;
  }
  $\hat\Delta \gets \hat\theta_p^{(t)} - \hat\theta_p^{(c)}$;\ \ \textbf{CS:} $\hat\Delta \pm \sqrt{h_c^2 + h_t^2}$\;
  \lIf{$|\hat\Delta| > \sqrt{h_c^2 + h_t^2}$}{\textbf{reject} $H_0 : Q_t(p) = Q_c(p)$}
}
\caption{Two-arm, clustered, asymptotically anytime-valid quantile treatment effect test (CAQ)}
\label{alg:twoarm}
\end{algorithm}

\section{3. Results}\label{sec-results}

\textbf{Simulation design.} All runs use \(p = 0.90\),
\(\alpha = 0.05\), and cluster sizes
\(N_i \sim \mathrm{Uniform}\{5, \dots, 15\}\) (mean \(\approx 10\)). A
normal DGP (data-generating process) places tunable intra-cluster
correlation
\(\mathrm{ICC} = \sigma_Z^2/(\sigma_Z^2 + \sigma_\varepsilon^2)\) with
total marginal variance fixed at \(2\) (so \(Q(p) \approx 1.8124\) is
held fixed as ICC varies), and a heavy-tailed lognormal DGP stresses the
tails. Two-arm effects are set by scaling the treatment arm's standard
deviation to a target \(\Delta_p\). We report, one-arm, the
\textbf{cumulative miscoverage rate} (a run fails if its CS ever misses
\(Q(p)\)) and, two-arm, the \textbf{rejection rate} (FPR under the null,
power otherwise). Full design details in \hyperref[app-b]{Appendix B}.
Throughout, \(M = 500\) simulations run over \(K = 50\) to \(1000\)
units (balanced across arms in the two-arm case), and we compare CAQ
against a peeked cluster-robust fixed-horizon interval (Deng et
al.~\hyperref[ref1]{{[}1{]}}) and a cluster-ignoring sequential interval
(Howard-Ramdas \hyperref[ref2]{{[}2{]}}).

\textbf{Results (One-arm).} Our simulations in Table \ref{tab:coverage}
reveal a clear difference in how well each method controls the
cumulative miscoverage rate: the baselines fail for distinct reasons,
while CAQ remains robust. Peeking inflates error regardless of
clustering (Deng et al.), while ignoring clustering inflates error as
intra-cluster correlation (ICC) grows (Howard-Ramdas). Only CAQ reliably
controls miscoverage across all regimes, and it does so at a modest cost
in width. Its intervals widen only as much as the clustering requires,
and when clustering is absent, they are even narrower than
Howard-Ramdas's i.i.d. width (\hyperref[app-c]{Appendix C}).

\begin{table}[H]
\centering
\resizebox{\textwidth}{!}{%
\begin{tabular}{llccc}
\hline
Distribution & ICC & Lu-Oh 2026 (CAQ) & Deng et al.\ 2018 (peeked) & Howard-Ramdas 2022 (ignore cluster) \\
\hline
Normal    & 0.00 & 0.026 & 0.324 & 0.000 \\
Normal    & 0.50 & 0.030 & 0.358 & 0.098 \\
Normal    & 0.90 & 0.032 & 0.360 & 0.448 \\
Lognormal & 0.50 & 0.036 & 0.376 & 0.218 \\
\hline
\end{tabular}%
}
\caption{One-arm cumulative miscoverage rate at representative ICC values (target $\alpha = 0.05$). CAQ stays at or below $\alpha$ throughout; ignoring clustering (Howard-Ramdas) worsens as ICC grows, while peeking (Deng et al.) fails throughout.}
\label{tab:coverage}
\end{table}

\begin{figure}[H]
\centering
\includegraphics[width=0.85\textwidth]{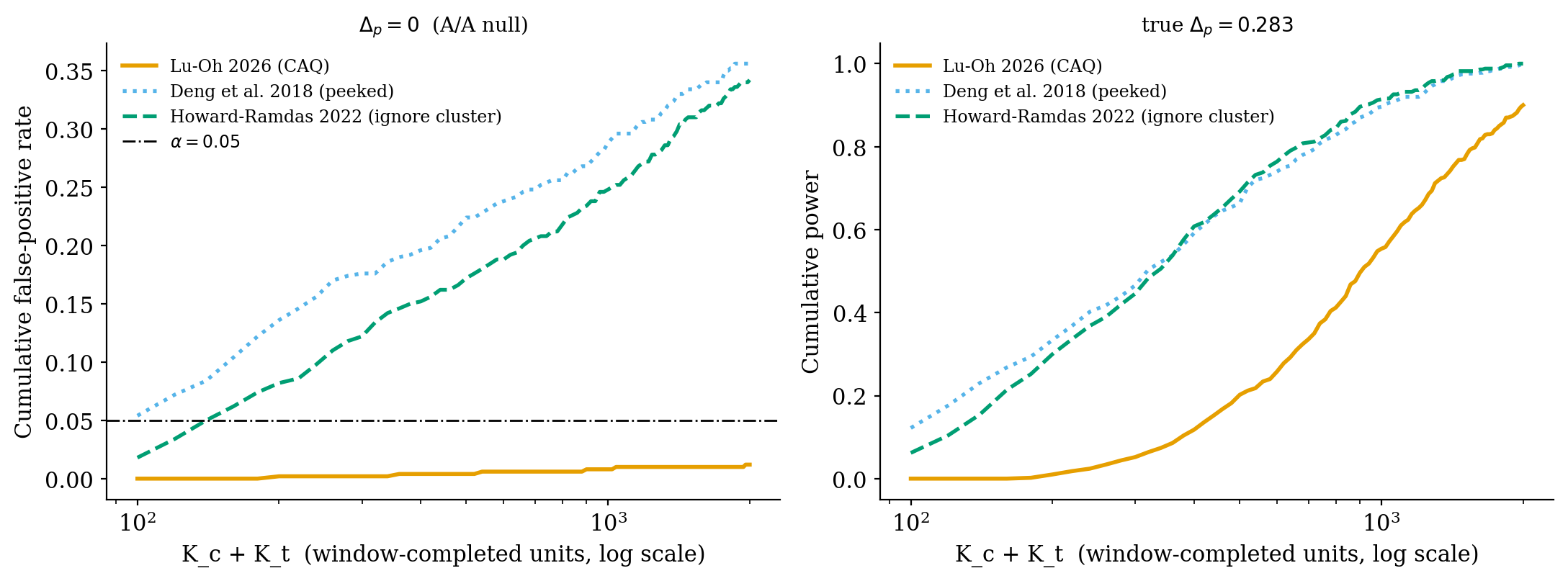}
\caption{Two-arm QTE test across the sequential looks (ICC $= 0.5$). \textbf{Left:} cumulative false-positive rate under the A/A null; only CAQ stays $\le \alpha = 0.05$. \textbf{Right:} cumulative power at $\Delta_p = 0.283$; Deng and Howard-Ramdas sit higher but are FPR-invalid, so not comparable.}
\label{fig:twoarm}
\end{figure}

\textbf{Results (Two-arm).} Crucially, CAQ is the only method that
remains both sequential and valid (Figure \ref{fig:twoarm}), holding the
false-positive rate at \(0.012\) (target \(\le \alpha\)). Because the
baselines severely inflate the FPR under continuous monitoring and
clustering, their apparent power advantages cannot be meaningfully
compared. However, for larger true quantile differences (Table
\ref{tab:fprpower}) or as the sample size increases (Figure
\ref{fig:twoarm}), CAQ achieves power comparable to the baselines.

\begin{table}[H]
\centering
\resizebox{\textwidth}{!}{%
\begin{tabular}{lccc}
\hline
Metric & Lu-Oh 2026 (CAQ) & Deng et al.\ 2018 (peeked) & Howard-Ramdas 2022 (ignore cluster) \\
\hline
FPR (target $\le 0.05$) & 0.012 & 0.356 & 0.342 \\
Power @ 0.141           & 0.286 & 0.774 & 0.760 \\
Power @ 0.283           & 0.900 & 0.998 & 1.000 \\
Power @ 0.424           & 1.000 & 1.000 & 1.000 \\
\hline
\end{tabular}%
}
\caption{Two-arm cumulative FPR and power. Only CAQ controls FPR at $\le \alpha = 0.05$; the peeked (Deng et al.) and cluster-ignoring (Howard-Ramdas) tests inflate FPR, so their higher power is not comparable.}
\label{tab:fprpower}
\end{table}

\section{4. Discussion}\label{discussion}

To our knowledge CAQ is the first cluster-robust, asymptotically
anytime-valid method for a \emph{quantile}, and hence for the quantile
treatment effect, closing the gap left open by prior work that only
handled pairs of the three axes. We prove CAQ is a
\((1-\alpha)\)-AsympCS for both the clustered quantile and the QTE, and
our simulations confirm it is the only method that controls both one-arm
miscoverage and two-arm false-positive rate under clustering and
continuous monitoring while still detecting effects as they accumulate.

\section{References}\label{references}

\begin{enumerate}
\def\labelenumi{\arabic{enumi}.}
\tightlist
\item
  \protect\phantomsection\label{ref1}{}Deng, Knoblich \& Lu (2018).
  Applying the delta method in metric analytics. KDD '18.
  arXiv:1803.06336.
\item
  \protect\phantomsection\label{ref2}{}Howard \& Ramdas (2022).
  Sequential estimation of quantiles with applications to A/B testing
  and best-arm identification. Bernoulli. arXiv:1906.09712.
\item
  \protect\phantomsection\label{ref3}{}Lindon, Sanden \& Shirikian
  (2022). Rapid regression detection in software deployments through
  sequential testing. KDD '22. arXiv:2205.14762.
\item
  \protect\phantomsection\label{ref4}{}Liu et al.~(2019). Large-scale
  online experimentation with quantile metrics. JSM Proceedings.
  arXiv:1903.08762.
\item
  \protect\phantomsection\label{ref5}{}Wang \& Zhang (2021). CONQ:
  continuous quantile treatment effects for large-scale online
  controlled experiments. WSDM '21. arXiv:2010.15326.
\item
  \protect\phantomsection\label{ref6}{}Waudby-Smith et al.~(2024).
  Time-uniform central limit theory and asymptotic confidence sequences.
  Ann. Statist. arXiv:2103.06476.
\end{enumerate}

\section{Appendix A: Method derivation}\label{app-a}

\emph{Standing assumptions.} Per arm, units enter i.i.d. with
\(\mu_N = \mathbb E[N_i] \in (0, \infty)\) and
\(\mathbb E[N_i^2] < \infty\); \(F\) is twice continuously
differentiable near \(Q(p)\) with density \(f = F'(Q(p)) > 0\); and the
per-unit influence function \(\phi_i\) (defined below) has variance
\(V = \operatorname{var}(\phi_i) \in (0, \infty)\). For the two-arm
test, assignment is randomized independently of outcomes, with known,
fixed assignment probabilities \(\pi_c, \pi_t \in (0, 1)\).

\textbf{One-arm.} A first-order Taylor expansion of the ratio \(s/n\)
about \((\mu_S, \mu_N)\) writes \(\hat p\) as an average of i.i.d.
per-unit terms plus a remainder,

\[\hat p - p = \frac1K\sum_{i=1}^K \phi_i + R_K, \qquad \phi_i = \frac{S_i - p\,N_i}{\mu_N}, \qquad \mathbb{E}[\phi_i] = 0,\]

with cluster-robust influence variance \(\operatorname{var}(\phi_i)\)
estimated by
\(\hat V = \bar N^{-2}\big(\hat\sigma_S^2 - 2\hat p\,\hat\sigma_{SN} + \hat p^2\hat\sigma_N^2\big)\),
the sample variances and covariance of \(\{S_i\}\) and \(\{N_i\}\).

The remainder is exactly
\(R_K = (\bar S - p\,\bar N)(\mu_N - \bar N)/(\mu_N \cdot \bar N)\).
This is a product of two mean-zero i.i.d. averages,
\(\bar S - p\,\bar N = \frac1K\sum_i(S_i - pN_i)\) and
\(\mu_N - \bar N = -\frac1K\sum_i(N_i - \mu_N)\). By the law of the
iterated logarithm \hyperref[refD]{{[}D{]}}, each is
\(O(\sqrt{\log\log K/K})\) a.s. The strong law of large numbers (SLLN)
sends the denominator \(\mu_N\cdot\bar N \to \mu_N^2 > 0\) a.s. So
\(R_K = O(\log\log K/K) = o(\sqrt{\log K/K})\), the rate that makes
\(\hat p\) an asymptotically linear time-uniform estimator of \(p\)
\hyperref[ref6]{{[}6{]}}. Such an estimator inherits a
\((1-\alpha)\)-AsympCS \hyperref[ref6]{{[}6{]}},

\[\hat p \pm \beta_K, \qquad \beta_K = b(K)\sqrt{\hat V / K}, \qquad b(K) = \sqrt{\frac{2(K\rho^2+1)}{K\rho^2}\,\log\!\frac{\sqrt{K\rho^2+1}}{\alpha}}.\]

Its coverage is time-uniform but asymptotic, holding at all \(K\) at
once in the large-sample limit, not exactly at finite \(K\);
\(\rho > 0\) tunes where the sequence is tightest, not its validity.

This band is for the CDF value \(p\), not the quantile \(Q(p)\). Because
\(F\) is monotone, we can invert it exactly into the quantile confidence
sequence below:

\[\big[\,X_{(L)},\ X_{(U)}\,\big], \quad L = \lceil n(p-\beta_K)\rceil, \ \ U = \lfloor n(p+\beta_K)\rfloor + 1, \qquad h := \tfrac12\big(X_{(U)} - X_{(L)}\big).\]

This exact inversion preserves the time-uniform guarantee. Thus, it
yields a \((1-\alpha)\)-AsympCS for the clustered quantile \(Q(p)\) that
is valid under any number of sequential looks.

\textbf{Two-arm.} The two-arm CS \(\hat\Delta \pm \sqrt{h_c^2 + h_t^2}\)
(Algorithm \ref{alg:twoarm}) is not two intervals glued together but a
\emph{single} application of the same AsympCS result
\hyperref[ref6]{{[}6{]}} to one estimator over the total count
\(m = K_c + K_t\). The argument is four steps.

\emph{Step 1: per-arm quantile linearization (clustered Bahadur
representation).} The classical Bahadur representation
\hyperref[refB]{{[}B{]}} assumes i.i.d. observations; here the ranked
events are within-unit correlated, so we rebuild it leveraging the
i.i.d.-across-units structure.

Work in one arm, drop the superscript \(a\), and let \(\hat F_K\) denote
the pooled empirical CDF (so the one-arm estimate is
\(\hat p = \hat F_K(Q(p))\)). Decompose the empirical CDF at the sample
quantile,

\[\hat F_K(\hat\theta_p) = F(\hat\theta_p) + [\hat F_K - F](\hat\theta_p),\]

and read off each piece.

\begin{enumerate}
\def\labelenumi{(\roman{enumi})}
\item
  \emph{Left-hand side.} \(\hat F_K(\hat\theta_p) = p + O(1/n)\),
  because \(\hat\theta_p = X_{(\lceil np\rceil)}\) is the smallest event
  whose empirical CDF reaches \(p\) and \(\hat F_K\) jumps by \(1/n\) at
  each of the \(n\) pooled events, so
  \(\hat F_K(\hat\theta_p) = \lceil np\rceil / n\).
\item
  \emph{First term on the right.} By a first-order Taylor expansion of
  the true CDF at \(Q(p)\),
  \(F(\hat\theta_p) = p + f\,(\hat\theta_p - Q(p)) + O\big((\hat\theta_p - Q(p))^2\big)\).
  The remainder is
  \(O\big((\hat\theta_p - Q(p))^2\big) = O(\log\log K/K)\) a.s. (by the
  empirical-process law of the iterated logarithm
  \hyperref[refA]{{[}A{]}}), negligible next to the \(\sqrt{\log K/K}\)
  boundary.
\item
  \emph{Second term on the right.}
  \([\hat F_K - F](\hat\theta_p) = [\hat F_K(Q(p)) - p] + o(\sqrt{\log
  K/K}) = (\hat p - p) + o(\sqrt{\log K/K})\) a.s., by the oscillation
  modulus for the clustered empirical process \hyperref[refC]{{[}C{]}}.
\end{enumerate}

Putting (i)-(iii) together and canceling \(p\),

\[p + O(1/n) = p + f\,(\hat\theta_p - Q(p)) + (\hat p - p) + o\!\left(\sqrt{\tfrac{\log K}{K}}\right) \ \Longrightarrow\ f\,(\hat\theta_p - Q(p)) = -(\hat p - p) + o\!\left(\sqrt{\tfrac{\log K}{K}}\right).\]

Dividing by \(f\) and substituting the one-arm CDF linearization
\(\hat p - p = \frac1K\sum_i \phi_i + R_K\) from above gives, per arm,

\[\hat\theta_p^{(a)} - Q_a(p) = -\frac{\hat p^{(a)} - p}{f_a} + o(\cdot) = \frac1{K_a}\sum_{i=1}^{K_a}\psi_i^{(a)} + o\!\left(\sqrt{\tfrac{\log K_a}{K_a}}\right)\ \text{a.s.}, \qquad \psi_i^{(a)} = -\frac{\phi_i^{(a)}}{f_a}, \quad \operatorname{var}(\psi_i^{(a)}) = \frac{V_a}{f_a^2}.\]

This is the clustered Bahadur representation.

\emph{Step 2: fuse the two arms into one estimator over \(m\).} Step 1
leaves two averages over different index sets; inverse-probability
weighting (IPW) collapses them onto the single total count \(m\). Model
the experiment as a single i.i.d. arrival stream
\(W_i = (A_i, \mathbf X_i)\), \(i = 1, \dots, m\), where \(A_i \sim
\mathrm{Bernoulli}(\pi_t)\) is the arm assignment, \(\mathbf X_i\) the
unit's events, and \(\pi_a = K_a/m\) the assignment probability (the
arm's traffic allocation). The IPW per-arrival influence function

\[\varphi(W_i) = \frac{A_i}{\pi_t}\,\psi^{(t)}(W_i) - \frac{1 - A_i}{\pi_c}\,\psi^{(c)}(W_i), \qquad \mathbb E[\varphi] = 0,\]

has running average
\(\frac1m\sum_{i=1}^m \varphi(W_i) = \frac1{K_t}\sum_{\text{treatment}}\psi^{(t)} -
\frac1{K_c}\sum_{\text{control}}\psi^{(c)}\) (exact, since \(\pi_a\) is
known and constant), and because each arrival loads exactly one arm
there is no cross term:

\[\operatorname{var}(\varphi) = \frac{V_t/f_t^2}{\pi_t} + \frac{V_c/f_c^2}{\pi_c} =: \Sigma^2 \in (0, \infty).\]

Linearity of \(x-y\) and exact reweighting add no new remainder, so

\[\hat\Delta - \Delta_p = \frac1m\sum_{i=1}^m \varphi(W_i) + o\!\left(\sqrt{\tfrac{\log m}{m}}\right)\ \text{a.s.}\]

Thus \(\hat\Delta\) is a \emph{single} asymptotically linear
time-uniform estimator of \(\Delta_p\) over the total count \(m\).

\emph{Step 3: form the AsympCS.} The AsympCS result
\hyperref[ref6]{{[}6{]}} at index \(m\) then yields the
\((1-\alpha)\)-AsympCS

\[\hat\Delta \pm b(m)\sqrt{\Sigma^2/m}, \qquad \frac{\Sigma^2}{m} = \mathrm{SE}_c^2 + \mathrm{SE}_t^2, \quad \mathrm{SE}_a = \frac{\sqrt{V_a/K_a}}{f_a},\]

\emph{Step 4: make it density-free.} Since
\(X^{(a)}_{(r)} = \hat F_a^{-1}(r/n^{(a)})\), the half-width can be
expressed as

\[h_a = \tfrac12\big[\hat F_a^{-1}(p+\beta_a) - \hat F_a^{-1}(p-\beta_a)\big] + O(1/n^{(a)}),\]

with \(\beta_a = b(m)\sqrt{\hat V_a/K_a}\). Then by the mean value
theorem, with \((F_a^{-1})'(p) = 1/f_a\),

\[h_a = \tfrac12\cdot 2\beta_a\cdot\frac{1}{f_a}\,(1+o(1)) + O(1/n^{(a)}) = b(m)\,\frac{\sqrt{\hat V_a/K_a}}{f_a} + o(\beta_a) = b(m)\,\mathrm{SE}_a + o(\sqrt{\log m/m}).\]

The leading term \(b(m)\,\mathrm{SE}_a\) is
\(\Theta(\sqrt{\log m})\cdot\Theta(1/\sqrt m) = \Theta(\sqrt{\log m/m})\),
so \(h_a = b(m)\,\mathrm{SE}_a(1+o(1))\). Hence,

\[\sqrt{h_c^2 + h_t^2} = b(m)\sqrt{\mathrm{SE}_c^2 + \mathrm{SE}_t^2}\,\big(1 + o(1)\big)\ \text{a.s.},\]

so \(\sqrt{h_c^2 + h_t^2}\) is asymptotically equal to the Step 3
half-width \(b(m)\sqrt{\Sigma^2/m}\) (recall
\(\Sigma^2/m = \mathrm{SE}_c^2 + \mathrm{SE}_t^2\)). Thus,
\(\hat\Delta \pm \sqrt{h_c^2 + h_t^2}\) is a \((1-\alpha)\)-AsympCS for
\(\Delta_p\), valid at every look \(m\) and for any allocation
\((K_c, K_t)\).

\section{Appendix B: Simulation details}\label{app-b}

\emph{Data-generating process.} For unit \(i\) and within-unit event
\(j = 1, \dots, N_i\), the \textbf{normal DGP} draws
\(Z_i \sim \mathcal N(0, \sigma_Z^2)\) and
\(\varepsilon_{ij} \sim \mathcal N(0, \sigma_\varepsilon^2)\) and sets
\(X_{ij} = Z_i + \varepsilon_{ij}\), so each event is marginally
\(\mathcal N(0, \sigma_Z^2 + \sigma_\varepsilon^2)\). The shared \(Z_i\)
is what correlates a unit's events, with intra-cluster correlation
\[\mathrm{ICC} = \frac{\sigma_Z^2}{\sigma_Z^2 + \sigma_\varepsilon^2}.\]
To isolate the effect of clustering we hold the total marginal variance
fixed at \(\sigma_Z^2 + \sigma_\varepsilon^2 = 2\) (so the marginal
stays \(\mathcal N(0, 2)\) and the true quantile \(Q(p) \approx 1.8124\)
does not move) and only shift variance between the unit-level
(\(\sigma_Z^2\)) and event-level (\(\sigma_\varepsilon^2\)) components,
\[\sigma_Z^2 = 2\,\mathrm{ICC}, \qquad \sigma_\varepsilon^2 = 2\,(1 - \mathrm{ICC}),\]
so \(\mathrm{ICC} = 0\) (\(\sigma_Z = 0\)) is i.i.d. events and larger
\(\mathrm{ICC}\) is stronger clustering. The \textbf{heavy-tailed DGP}
instead sets \(X_{ij} = e^{Z_i} + e^{\varepsilon_{ij}}\), a sum of two
lognormals; its marginal quantile has no closed form and is estimated by
Monte Carlo.

All two-arm runs use the normal DGP: with
\(\sigma_{\mathrm{tot}} = \sqrt{\sigma_Z^2 + \sigma_\varepsilon^2}\) the
marginal \(p\)-quantile is \(Q(p) = \sigma_{\mathrm{tot}}\,z_p\)
(\(z_p = \Phi^{-1}(p)\)), and scaling the treatment arm's SD by
\(k = 1 + \Delta_p / (\sigma_{\mathrm{tot}}^{(c)}\,z_p)\) gives
\(Q_t(p) - Q_c(p) = \Delta_p\).

\emph{Performance metrics.} Over \(M\) independent runs, at each look
\(K\) a one-arm method returns a confidence sequence \(\mathrm{CI}_K\);
the \textbf{cumulative miscoverage rate} is
\[\widehat{\mathrm{Miscoverage}} = \frac1M \sum_{m=1}^M \mathbf 1\Big\{\, \exists\,K : Q(p) \notin \mathrm{CI}_K^{(m)} \,\Big\} = 1 - \widehat{\mathrm{Coverage}},\]
the fraction of runs whose sequence misses the truth at some look up to
\(K\). A \emph{nonasymptotic} time-uniform CS satisfies
\(\mathbb P(\exists K : Q(p) \notin \mathrm{CI}_K) \le \alpha\) exactly;
CAQ is an \emph{asymptotic} CS (AsympCS \hyperref[ref6]{{[}6{]}}),
guaranteed asymptotically equivalent to one that does, and we report
empirical miscoverage to gauge how well this holds. For the two-arm test
a run counts as a rejection if it declares significance at any look,
\[\widehat{\mathrm{Rej}} = \frac1M \sum_{m=1}^M \mathbf 1\big\{\, \exists\,K : \text{run } m \text{ rejects at look } K \,\big\},\]
the \textbf{false-positive rate (FPR)} under \(\Delta_p = 0\) and the
\textbf{power} under \(\Delta_p > 0\).

\section{Appendix C: One-arm validity and interval width}\label{app-c}

\begin{figure}[H]
\centering
\includegraphics[width=\textwidth]{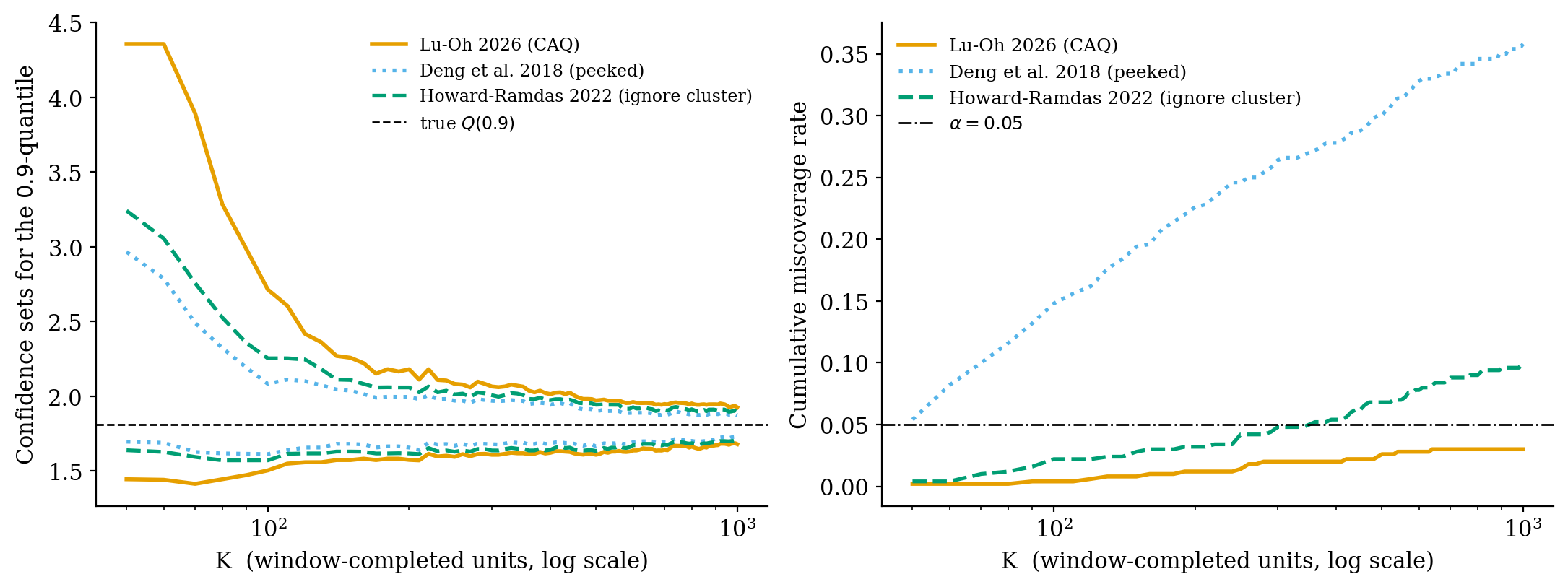}
\caption{Clustered $0.9$-quantile validity at ICC $= 0.5$ (same setup as Table \ref{tab:coverage}). \textbf{Left:} single-run confidence sets narrowing to the truth (dashed); CAQ is the widest, guarding against the clustered design effect. \textbf{Right:} cumulative miscoverage rate: CAQ stays under $\alpha = 0.05$ at every look, while Deng (peeking) and Howard-Ramdas (ignoring clustering) exceed it.}
\label{fig:flatk}
\end{figure}

\begin{figure}[H]
\centering
\includegraphics[width=\textwidth]{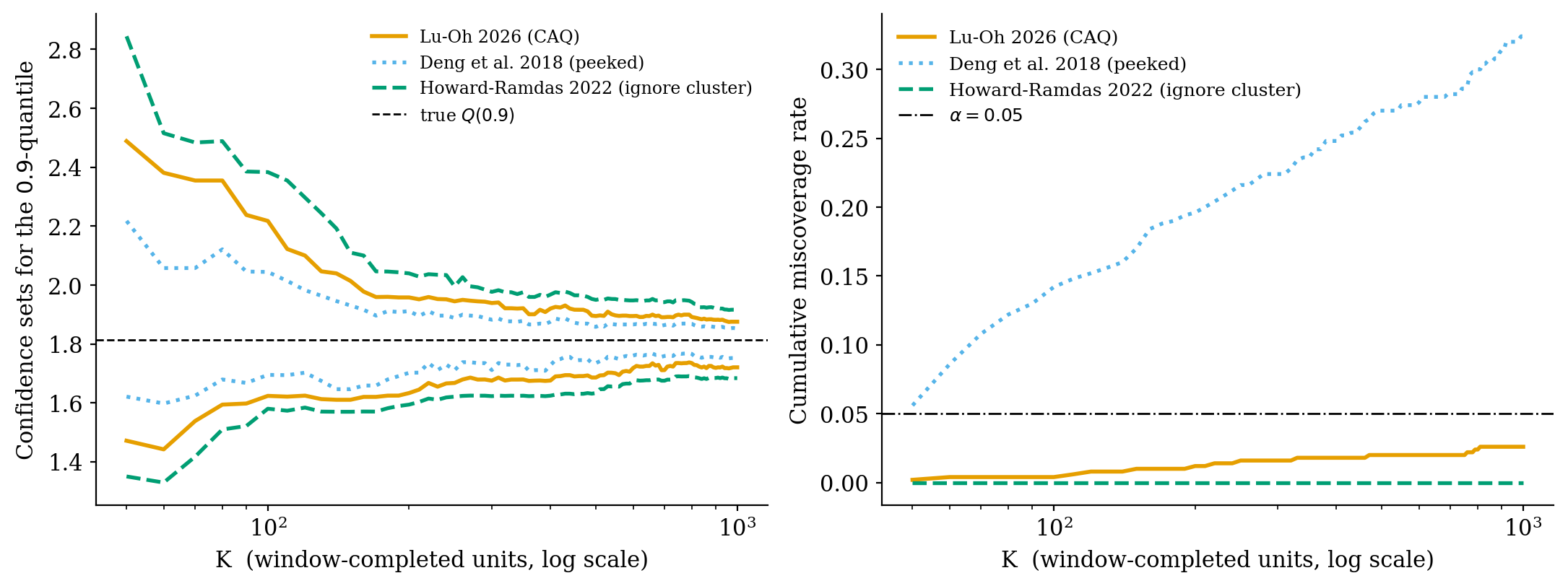}
\caption{ICC $= 0$ case (i.i.d.\ events; same setup as Figure \ref{fig:flatk} with $\sigma_Z = 0$), isolating the two failure modes. \textbf{Left:} Howard-Ramdas now has the widest confidence sets, showing CAQ's width is adaptive. \textbf{Right:} with no clustering to ignore, both CAQ and Howard-Ramdas's miscoverage stay $\le \alpha = 0.05$ while Deng still fails ($\approx 0.32$), pinning its error to peeking.}
\label{fig:icc0}
\end{figure}

CAQ's validity costs interval width, but adaptively: its boundary is
sized from the estimated cluster-robust variance \(\hat V\), so it is
widest under strong clustering (Figure \ref{fig:flatk}) yet shrinks to
near the i.i.d. width when ICC \(\approx 0\), where it is in fact
\emph{narrower} than Howard-Ramdas (Figure \ref{fig:icc0}). The
difference is the guarantee: Howard-Ramdas is nonasymptotic and
distribution-free, so its boundary is conservative and worst-case wide,
whereas CAQ claims only asymptotic validity.

\section{Appendix references}\label{appendix-references}

\begin{enumerate}
\def\labelenumi{\Alph{enumi}.}
\tightlist
\item
  \protect\phantomsection\label{refA}{}Alexander \& Talagrand (1989).
  The law of the iterated logarithm for empirical processes on
  Vapnik-Chervonenkis classes. J. Multivariate Anal. 30(2), 155-166.
\item
  \protect\phantomsection\label{refB}{}Bahadur (1966). A note on
  quantiles in large samples. Ann. Math. Statist. 37(3), 577-580.
\item
  \protect\phantomsection\label{refC}{}Gine \& Koltchinskii (2006).
  Concentration inequalities and asymptotic results for ratio type
  empirical processes. Ann. Probab. 34(3), 1143-1216.
\item
  \protect\phantomsection\label{refD}{}Hartman \& Wintner (1941). On the
  law of the iterated logarithm. Amer. J. Math. 63(1), 169-176.
\end{enumerate}

\end{document}